\documentclass[aps,prb,twocolumn,superscriptaddress]{revtex4-2}

\usepackage{amsmath, amssymb,  graphicx}

\usepackage[colorlinks=true ,urlcolor=blue,urlbordercolor={0 1 1}]{hyperref}

\usepackage{color}
\usepackage{xcolor}
\usepackage{braket}

\usepackage[utf8]{inputenc}
\usepackage{bm}
\usepackage{verbatim}
\usepackage{graphicx}
\usepackage{amsmath}
\usepackage{color}
\usepackage{float}
\usepackage{bbm}
\usepackage{amssymb}
\usepackage{slashed}
\usepackage{wasysym,bm,bbm,dsfont}

\begin{document}

\title{Holon metal, charge-density-wave and chiral superconductor from doping   fractional Chern insulator and SU(3)$_1$ chiral spin liquid}

\author{Ya-Hui Zhang}
\affiliation{Department of Physics and Astronomy, Johns Hopkins University, Baltimore, Maryland 21218, USA}

\date{\today}

\begin{abstract}
Recent experiments have observed superconductivity proximate to the $\nu = -2/3$ fractional quantum anomalous Hall (FQAH) insulator in twisted MoTe$_2$. A critical open question is whether the underlying normal state is a Fermi liquid with a large Fermi surface or a strongly correlated metal with low carrier density. In this work, we develop a theory of the phases emerging from doping the $\nu = -2/3$ fractional Chern insulator (FCI). We establish a duality between this problem and doping a gapped SU(3)$_1$ chiral spin liquid (CSL). In both scenarios, one possible metallic  state upon doping is a holon metal featuring three small Fermi pockets. These pockets are formed by spinless charge $-e$ holons in the CSL case and charge $-e/3$ fractionalized holes in the FCI case. While the holon metal is susceptible to pairing instabilities driven by gauge fluctuations—leading to a charge density wave (CDW) metal—it may be stabilized by a magnetic field, where it would exhibit an anomalous quantum oscillation period. We identify two distinct chiral superconducting phases that can emerge either from the CDW metal or directly from the holon metal. Finally, we argue that superconductivity arising directly from an anyon gas is not likely if the lowest-energy anyon carries the elementary charge $e/3$.\end{abstract}

\maketitle

\section{Introduction}

Fractional phases in the topological  bands of moir\'e systems\cite{zhang2019nearly,wu2019topological} have recently attracted many attentions.  At fractional filling of a flat Chern band, fractional Chern insulator (FCI) is expected\cite{sun2011nearly,sheng2011fractional,neupert2011fractional,wang2011fractional,tang2011high,regnault2011fractional,bergholtz2013topological,parameswaran2013fractional,PhysRevB.84.165107}. Indeed FCI states with fractional quantum anomalous Hall effects were observed in twisted MoTe$_2$ homobilayer \cite{cai2023signatures,zeng2023integer,park2023observation,PhysRevX.13.031037} and in rhombohedrally stacked multilayer graphene aligned with hBN\cite{2023arXiv230917436L}.  These exciting experimental progresses  already initiated many theoretical studies \cite{wu2019topological,yu2020giant,devakul2021magic,li2021spontaneous,crepel2023fci,wang2023fractional,reddy2023fractional,2023arXiv230809697X,2023arXiv230914429Y,PhysRevLett.131.136501,PhysRevLett.131.136502,morales2023magic,song2023phase,dong2023theory,zhou2023fractional,dong2023anomalous,guo2023theory,kwan2023moir}. The major focus is on the incompressible phase at the commensurate filling. On the other hand, possible compressible phases upon doping the fractional state have not been well explored. Theoretically there have been proposals for new types of superconductivity\cite{Laughlin1988PRL,Laughlin1988Science,HalperinAnyonSC1990,FisherLeeAnyonSC1991,tang2013superconductivity,kim2025topological,shi2024doping,darius2025doping} connected to the FCI state. More recently, signature of a superconductor nearby the $\nu=-\frac{2}{3}$ FCI in the twisted MoTe$_2$ system was reported experimentally\cite{xu2025signatures}. Motivated by the experimental observation, in this work we study possible compressible phases from doping the $\nu=-\frac{2}{3}$ FCI.  Our approach is different from the earlier works on anyon superconductivity\cite{Laughlin1988PRL,Laughlin1988Science,HalperinAnyonSC1990,FisherLeeAnyonSC1991,tang2013superconductivity,kim2025topological}. It shares certain similarity  to Ref.~\onlinecite{shi2024doping}, but also with some difference which we will clarify later.

We also propose to unify the problem of doping the FCI and another seemingly unrelated problem: doping an SU(3)$_1$ chiral spin liquid (CSL) in an SU(3) Hubbard model.  Chiral spin liquid\cite{kalmeyer1987equivalence,wen_1989}
is a state emerging in the Mott insulator at total filling $n=1$ of a Hubbard model. It  hosts a topological order equivalent to a fractional quantum Hall phase.  CSL has both a charge gap and a spin gap. Normally the charge gap is significantly larger than the spin gap deep inside the Mott regime. However, when doping the CSL or increasing the bandwidth, the charge gap can be closed while the spin gap remains finite. In this case, at the energy scale below the spin gap, an SU(2)$_1$ CSL can be shown to be equivalent to a $\nu=-\frac{1}{2}$ Laughlin state of the charge $2e$ Cooper pair\cite{song2022deconfined}. Therefore anyon superconductor from doping an SU(2)$_1$ CSL\cite{song2021doping,divic2024anyon} is dual to anyon superfluid from doping a $\nu=-\frac{1}{2}$ bosonic Laughlin state.  In this work we establish a similar duality between the SU(3)$_1$ CSL and the $\nu=-\frac{1}{3}$ fermionic Laughlin state.  Assuming the spin gap is not closed upon doping the CSL, the universal physics is  the same as doping the fermionic FCI, just now the single elctron in the doped FCI side is dual to a spin-singlet charge $3e$ trion in the doped CSL side.

To further demonstrate the duality, we derive the same low energy effective field theory for both problems. In the CSL side we use the standard slave boson theory and find the slave boson has three minima in its dispersion, as required by the projective translation symmetry. At low energy each minima provides a bosonic field $\varphi_I$, $I=1,2,3$, which feels an emergent internal flux and is at effective filling $\nu_m=-1$.  For the doped SU(2)$_1$ CSL, we only have two flavors and they can form a bosonic integer quantum Hall state, leading to a superconductor for the electron\cite{song2021doping}.  In contrast, now the only translation invariant state is to put each flavor $\varphi_I$ in a bosonic composite Fermi liquid (CFL) or Pfaffian state. Let us focus on the simplest CFL scenario. The final phase in terms of electron turns out to be a holon metal\cite{kaul2008algebraic}, which has three small Fermi pockets formed by spinless charge $-e$ holon.  In the dual side of doped FCI, we can formulate the same effective theory, where $\varphi_I$ is now the vortex of the composite boson in the standard theory of Laughlin state. In the end if we put each $\varphi_I$ in a bosonic CFL state, again we get an exotic metal with three small pocket, but now formed by charge $-e/3$ fractionalized hole.  For simplicity, we still dub the phase as holon metal. A similar phase was proposed in Ref.~\cite{shi2024doping} for doped FCI and was called `secondary CFL'. However, the property of the phase is quite different from a CFL: (1) the phase is compressible and well defined in a range of density, while the CFL phase is restricted to a specific doping; (2) The phase has a finite drude weight and the conductivity $\sigma_{xx}$ could be arbitrarily large in the clean limit. In contrast the CFL phase has vanishing $\sigma_{xx}$ in the clean limit.  For this reason we prefer the name holon metal. Our holon metal is a simpler version of the state in Ref.~\cite{shi2024doping}. Intuitively this exotic metal may be conveniently constructed in this way: we fractionalize an electron into three fractionalized fermions with charge $e/3$, and then put each fermion in a $C=-1$ Chern insulator plus a small hole pocket.  

The holon metals in both doped CSL and doped FCI are likely unstable due to gauge field fluctuations. The natural descendant is a CDW metal from inter-flavor pairing of two Fermi pockets in the holon metal. In the doped CSL side this CDW metal is still exotic as the remaining one fermi surface is formed by the spinless charge $3e$ trion. In the doped FCI case we just have one single Fermi surface formed by the physical hole on top of a CDW order which triples the unit cell. In the context of doping the $\nu=-\frac{2}{3}$ FCI in twisted MoTe$_2$, the CDW metal in our theory has $\sigma_{xy} \approx 0$, different from the one in Ref.~\cite{shi2024doping} which has $\sigma_{xy} \approx \frac{e^2}{h}$ due to a background integer quantum Hall effect. Because the CDW order here is from inter-flavor pairing of the holon metal, we propose to suppress CDW order and stablilize the holon metal by applying an external magnetic field.  We also predict that the holon metal has quantum oscillation with an exotic frequency because each Fermi pocket sees an effective magnetic field $B_{\mathrm{eff}}$ which is $1/3$ of the physical magnetic field $B$.

The most interesting question is whether superconductivity is possible from doping the FCI or SU(3) CSL. In our framework, we do not find a natural  way to enter a superconductor phase directly from an anyon gas if the cheapest anyon is the elementary one with $e/3$ charge. This is very different from doping the SU(2) CSL or the bosonic Laughlin state.  Although we do not have a good understanding of the microscopic energetics, we can imagine two different superconductors: (1) an ordinary $p\pm ip$ pairing of the single Fermi surface in the CDW metal; (2) a more exotic fractionalized superconductor from intra-flavor pairing\cite{shi2024doping} of the holon metal. The two scenarios may be distinguished by different phase diagrams under the magnetic field $B$. In the second scenario, after the superconductor is killed, the system enters the holon metal with a large $\sigma_{xy} \approx \frac{2}{3} \frac{e^2}{h}$. On the other hand, in the first scenario, there should be an intermediate CDW metal phase with $\sigma_{xy} \approx 0$.  

The remaining paper is organized in the following way. In Sec.~\ref{sec:su2_boson} we review the duality for anyon superconductor/superfluid in doped SU(2) CSL and bosonic Laughlin state. In Sec.~\ref{sec:holon_metal_csl} we discuss the holon metal and CDW phase from doping the SU(3)$_1$ CSL. A similar theory is presented in Sec.~\ref{sec:holon_metal_fci} for doping the $\nu=-\frac{2}{3}$ FCI and we also discuss the experimental signatures of the holon metal and the CDW phase.  We briefly discuss the implication of our theory in the context of the experiment in twisted MoTe$_2$\cite{xu2025signatures} in Sec.~\ref{sec:discussion} and then conclude in Sec.~\ref{sec:conclusion}.

\section{Doping SU(2)$_1$ CSL and bosonic Laughlin state: anyon superfluid\label{sec:su2_boson}}

In this section, we review the construction of anyon superconductor from doping a SU(2)$_1$ chiral spin liquid in a spin 1/2 model. The theory is also dual to doping a bosonic Laughlin state at $\nu=\frac{1}{2}$. 

\subsection{Doped SU(2)$_1$ CSL}

Consider an SU(2) Hubbard model. At filling $n=1$ (half filling per spin), we can have a chiral spin liquid. As this work is concerned with low-energy effective theory, we will not specify the microscopic interaction to stabilize the CSL.  The CSL phase is within the Mott insulator phase and is mostly convenient described by the slave boson construction: $c_{i;\sigma}=b_i f_{i;\sigma}$ with $b_i$ the slave boson and $f_{i;\sigma}$ the Abrikosov fermion. There is a U(1) gauge redundancy: $b_i \rightarrow b_i e^{i\theta_i}$ and $f_{i;\sigma}\rightarrow f_{i;\sigma} e^{-i\theta_i}$.  Therefore, there is an internal U(1) gauge field $a_\mu$. Meanwhile, let us also introduce the U(1) probing field $A_{c;\mu}$ which represents the electromagnetic field in the real world. In the end, $f$ couples to $a$ and $b$ couples to $A_c-a$.  At $n=1$, the slave boson $b$ is in a Mott insulator phase due to Hubbard U and is simply gapped. Then we are left with the spinon $f_\sigma$. In the CSL phase, $f_\sigma$ is put in a mean field ansatz with $\pi$ flux per unit cell and forms a Chern insulator with $C=1$ for each spin $\sigma=\uparrow,\downarrow$.

The low energy effective theory is described by
\begin{equation}
    \mathcal L_{\mathrm{SU(2)_1 CSL}}=\mathcal L_b[\varphi,A_c-a]+\frac{2}{4\pi} a da
\end{equation}
where $ada$ is the abbreviation of the Chern-Simons term $\epsilon_{\mu\nu \sigma} a_\mu \partial_\nu a_\sigma$. Here $\epsilon$ is the anti-symmetric tensor.

$\mathcal L_b$ describes the dynamics of the gapped slave boson, which is now represented by $\varphi\sim b^\dagger$:

\begin{align}
    \mathcal L_b[\varphi,-A_c+a]&=\varphi^*\left(i\partial_t+(-A_{c;0}+a_0)\right)\varphi +\Delta \varphi^* \varphi\notag \\ 
    &+\frac{\hbar^2}{2m^*}\varphi^*\left(-i\vec \nabla-(-\vec A_c+\vec a)\right)^2 \varphi
\end{align}
where $\Delta>0$ is the charge gap. $\varphi$ represents the gapped charge excitations, which have self statistics $\theta=-\frac{\pi}{2}$.

Next we dope the system so the electron is at total filling $n=1-x$. We have $n_b=n_f=1-x$ on average. Now naively the spinon $f$ can not stay in a gapped Chern insulator phase and the spin gap is closed. In this simple case we expect that the slave boson is condensed and we just have a simple Fermi liquid coexisting with charge-density-wave (CDW) with doubled unit cell\cite{song2021doping}. However, there is actually a possibility that the spin gap remains finite and the low energy physics is dominated by finite density of spinless holons $b$. We will focus on this scenario with a finite spin gap in this work. Because the spin is gapped, the low energy degree of freedom is actually dominated by the physical charge 2e Cooper pair $\Delta=c_{\uparrow} c_{\downarrow}$.

To keep the spinon $f_\sigma$ in a gapped Chern insulator phase, we need to have a finite internal flux $\frac{\langle da \rangle}{2\pi}=-\frac{x}{2}$ per unit cell. Here $da$ indicates $\partial_x a_y-\partial_y a_x$. With the dynamically generated internal flux, $f_\sigma$ stays in the Chern insulator phase following the Streda's formula. Then the low energy theory is entirely from the slave boson $b$. $b$ is in a Mott insulator at $n=1$. Now at filling $n=1-x$, we basically dope its holes $\varphi \sim b^\dagger$ with density $n_{\varphi}=x$. However, due to the projective translation symmetry, $b$ feels a $\pi$ flux per triangular unit cell and the magnetic translation symmetry guarantees two minima. If one of them is at momentum $\mathbf k_1=(0,0)$, the other one is at $\mathbf k_2=(\pi,0)$.  We label the low energy  operator of $\varphi$ around these two minima as $\varphi_1(t,\mathbf r)$ and $\varphi_2(t,\mathbf r)$.  Then the physical translation symmetry acts as: $T_x: \varphi_1 \rightarrow \varphi_1, \varphi_2 \rightarrow - \varphi_2$, $T_y: \varphi_1 \rightarrow \varphi_2, \varphi_2 \rightarrow \varphi_1$. Each of them then is at density $n=\frac{x}{2}$ per triangular unit cell.

\begin{align}
    \mathcal L&=\sum_{I=1,2}\varphi^*_I\left(i\partial_t+(-A_{c;0}+a_0)\right)\varphi_I +\mu \varphi^*_I \varphi_I\notag \\ 
    &+\frac{\hbar^2}{2m^*}\varphi^*_I\left(-i\vec \nabla-(-\vec A_c+\vec a)\right)^2 \varphi_I +\frac{2}{4\pi} a da
\end{align}
where the chemical potential $\mu$ is introduced to fix the average density $n_{\varphi_I}=\frac{x}{2}$ per unit cell. Note variation of $a_0$ gives us $\frac{\langle da \rangle}{2\pi}=-\frac{1}{2}\sum_I n_{\varphi_I}=-\frac{x}{2}$ as expected. At $x=0$, $\varphi_I$ does not feel any flux. But at finite $x$, the holon $\varphi_I$ feels an effective flux with effective filling $\nu=-1$ of the associated Landau level. 

It is convenient to shift the definition of $a_\mu$ by $a_\mu \rightarrow a_\mu-A_{c;\mu}$, so the Lagrangian can be rewritten as:

\begin{align}
    \mathcal L&=\sum_I \varphi^*_I \left(i\partial_t+a_0\right)\varphi_I +\mu \varphi^*_I \varphi_I+\frac{\hbar^2}{2m^*}\varphi^*_I\left(-i\vec \nabla-\vec a\right)^2 \varphi_I  \notag \\ &+\frac{2}{4\pi} a da-\frac{2}{2\pi}A_c da +\frac{2}{4\pi} A_c d A_c
    \label{eq:dope_su2_csl}
\end{align}

When $\mu<0$, $\varphi$ is gapped. Then we can see that the low energy effective theory is the same as that of the $\nu=-\frac{1}{2}$ Laughlin state of a charge $2e$ boson with an additional term $\frac{2}{4\pi} A_c d A_c$.  Basically, an SU(2)$_1$ CSL is equivalent to the Laughlin state of the Cooper pair with an additional integer quantum Hall (IQH) state with $C=2$. This is why theory theory of doping a CSL is closely related to that of doping a Laughlin state.

Now let us tune $\mu$ to dope the CSL so that $\langle n_{\varphi_I} \rangle=\frac{x}{2}$. From variation of $a_0$ we can find $\frac{\langle da \rangle}{2\pi}=-\frac{x}{2}$ as expected. Next we need to discuss the ground state with finite density of $\varphi_I$. Due to the effective magnetic flux, simple boson condensation is not an option. Instead, $\varphi$ can form a bosonic integer quantum Hall (bIQH) state with $C=-2$. In the end this provides an additional $\frac{2}{4\pi} a da$ contribution from the bIQH phase, which cancels the self Chern-Simons term in the above Lagrangian. In the end the low energy theory is:

\begin{equation}
    \mathcal L_{\mathrm{SC}}=-\frac{2}{4\pi} A_c da+\frac{2}{4\pi} A_c d A_c
\end{equation}

This describes a charge $2e$ superconductor. The spin part has the same response as the chiral spin liquid and thus topologically one can show the superconductor is equivalent to a spin-singlet $d+id$ superconductor\cite{song2021doping}. The essence of the construction is to make the parton $f_\sigma$ and slave boson $b$ in quantum Hall insulators with oppostite $\sigma_{xy}$, so that the final theory does not have self Chern-Simons term for the internal gauge field $a_\mu$. The internal flux of $a_\mu$ carries physical charge $2e$ and gives the superconductivity.

\subsection{Doped bosonic Laughlin state}

Here we consider the dual problem of doping the bosonic Laughlin state. Suppose we have a finite density of bosons with charge $e$ and it forms a $\nu=-\frac{1}{2}$ Laughlin state with $\sigma_{xy}=-\frac{e^2}{2h}$ at density $n=\frac{1}{2}$. Then we tune the density to be $n=\frac{1}{2}-x$ and ask what the ground state is. In  the Landau level with continuous translation symmetry, we usually only have an anyon gas. However, if the parent Laughlin state is realized as a fractional Chern insulator (FCI) in a lattice, there is a possibility to achieve a superfluid phase. The physics is actually exactly the same as that of doping SU(2)$_1$ CSL discussed in the previous subsection. The major change is that we should now view the charge $2e$ Cooper pair in the doped CSL case as our charge $e$ physical boson. So we simply replace $2A_c$ with $A$. We also need to remove the $\frac{2}{4\pi} A_c d A_c$ response in the doped CSL theory in Eq.~\ref{eq:dope_su2_csl}.

In the following we provide an independent derivation of the effective theory. The low energy theory of the doped Laughlin state is described by:

\begin{align}
    \mathcal L&=\varphi^*\left(i\partial_t+a_0\right)\varphi +\mu \varphi^* \varphi\notag \\ 
    &+\frac{\hbar^2}{2m^*}\varphi^*\left(-i\vec \nabla-\vec a\right)^2 \varphi +\frac{2}{4\pi} a da-\frac{1}{2\pi}A da 
    \label{eq:dope_bosonic_laughlin}
\end{align}

One may view $\varphi^\dagger$ as the creation operator of the anyon with charge $e/2$ in the standard effective theory of the Laughlin state. One may view $\varphi$ as the vortex of the composite boson.  When the physical boson is at filling $n=\frac{1}{2}-x$, $\varphi$ is at density $n_{\varphi}=2x$.  This is also in agreement with the doped CSL theory where the filling of the electron is $n=1-2x$ so the Cooper pair density is $\frac{1}{2}-x$.

The next step is exactly the same as the doped CSL case. $\varphi$ feels an effective magnetic flux from $da$ and is at magnetic filling $\nu=-2$, so it can form a bosonic IQH phase with $C=-2$. In the end the final theory is just:
\begin{equation}
    \mathcal L_{\mathrm{Anyon-SF}}=-\frac{1}{2\pi}A da
\end{equation}
which describes a superfluid phase of the physical boson. Note that the topological order of the Laughlin state is also killed.

\section{Holon metal from doped SU(3)$_1$ CSL\label{sec:holon_metal_csl}}

We now can discuss the major problem of this work: doping the SU(3)$_1$ CSL and the fermionic fractional Chern insulator at $\nu=-\frac{2}{3}$. Similarly to the previous section, these two problems are dual to each other. Hence we can focus on the doped CSL problem for now. We will see that now we need to deal with a fermionic model even below the spin gap, which is quite different from the bosonic model in the previous section. In this case we do not find a natural way to get an anyon superconductor directly, instead a holon metal seems more natural, though it is unstable at lower temperature.

Now we consider a SU(3) Hubbard model:

\begin{align}
    \mathcal H_{\mathrm{SU(3)}}=-t \sum_{\alpha=1,2,3}\sum_{{\langle ij \rangle}} c^\dagger_{i;\alpha}c_{j;\alpha}+\frac{U}{2} \sum_i n_i(n_i-1)+...
\end{align}
where $...$ includes other terms which may be necessary to stabilize a SU(3)$_1$ chiral spin liquid at integer filling $n=1$. We put the model on a triangular lattice with full translation symmetry. 

To describe the CSL, we use the slave boson construction $c_{i;a}=b_i  f_{i;a}$ with $a=1,2,3$ as the flavor index. $f_a$ is put into a $C=1$ insulator with a $\frac{2\pi}{3}$ flux per unit cell.  Similarly, the slave boson $b$ will feel a $-2\pi/3$ flux per unit cell. Its dispersion will have three minima guaranteed by the projective translation symmetry.  Next we dope the system so that the average density is $n_b=n_f=1-x$ per triangular unit cell. Similar to the SU(2) CSL case, $b$ and $f$ share an internal U(1) gauge field $a_\mu$. We assign the physical charge to the slave boson $b$. So in the end $b$ couples to $A_c-a$ and $f$ couples to $a$.

To keep $f$ in a $C=1$ Chern insulator, we need $\frac{\langle da \rangle}{2\pi}=-\frac{x}{3}$.  Then the low energy theory is dominated by the slave boson $b$. Again $b$ feels an effective flux and its hole $\varphi\sim b^\dagger$ has three minima at momentum $\mathbf k_1=(0,0), \mathbf k_2=(\frac{2\pi}{3},0), \mathbf k_3=(\frac{4\pi}{3},0)$. We define the corresponding low energy field operator as $\varphi_I(t,x)$ with $I=1,2,3$. The physical translation symmetry acts as $T_x: \varphi_I \rightarrow  e^{i \frac{2\pi}{3} I} \varphi_I$ and $T_y: \varphi_1 \rightarrow \varphi_2, \varphi_2 \rightarrow \varphi_3, \varphi_3 \rightarrow \varphi_1$.  The low-energy effective theory is:

\begin{align}
    \mathcal L&=\sum_{I=1,2,3}\varphi^*_I\left(i\partial_t+(-A_{c;0}+a_0)\right)\varphi_I +\mu \varphi^*_I \varphi_I\notag \\ 
    &+\frac{\hbar^2}{2m^*}\varphi^*_I\left(-i\vec \nabla-(-\vec A_c+\vec a)\right)^2 \varphi_I +\frac{3}{4\pi} a da
\end{align}
where the chemical potential $\mu$ is introduced to fix the average density $n_{\varphi_I}=\frac{x}{3}$ per triangular unit cell. Note that the variation of $a_0$ gives $\frac{\langle da \rangle}{2\pi}=-\frac{1}{3}\sum_I n_{\varphi_I}=-\frac{x}{3}$ as expected. At finite $x$, the holon $\varphi_I$ feels an effective flux with effective filling $\nu=-1$ of the associated Landau level. 

We can again shift $a_\mu$ by $a_\mu \rightarrow a_\mu+A_{c;\mu}$, and rewrite the Lagrangian as:

\begin{align}
    \mathcal L&=\sum_I \varphi^*_I \left(i\partial_t+a_0\right)\varphi_I +\mu \varphi^*_I \varphi_I\notag \\ 
    &~~~+\frac{\hbar^2}{2m^*}\sum_I \varphi^*_I\left(-i\vec \nabla-\vec a\right)^2 \varphi_I \notag \\
    &~~~+\frac{3}{4\pi} a da+\frac{3}{2\pi}A_c da +\frac{3}{4\pi} A_c d A_c
    \label{eq:dope_su3_csl}
\end{align}

We can see that the theory of doped SU(3)$_1$ CSL is quite similar to doping a $\nu=-\frac{1}{3}$ Laughlin state of a charge 3e spin-singlet trion $C_{\mathrm{trion}}=c_1 c_2 c_3$, stacked with a $C=3$ integer quantum Hall (IQH) phase.  Thus the theory here can be directly applied to the doped FCI case with simple change.

At physical density $n_c=1-x$, we have $n_{\varphi_I}=\frac{x}{3}$. Meanwhile from variation of $a_0$ we find an average internal flux $\frac{\langle da \rangle}{2\pi}=-\frac{x}{3}$. So each $\varphi_I$ is again at effective filling $\nu_m=-1$ of the associated Landau level.  But now we have three flavors, so a simple gapped bosonic IQH phase is not possible.  If we restrict to translation symmetric phase, then we need to put each $\varphi_I$ form a bosonic CFL or Pfaffian phase given that its magnetic filling is $\nu_m=-1$. Let us focus on the scenario of CFL here. We will show that the final phase is an exotic metallic phase for the physical electron.

For each $\varphi_I$, we now put it into a bosonic CFL phase described by the Lagrangian:

\begin{equation}
    \mathcal L_{\varphi_I}=\mathcal L_{FS}[\psi_I, \alpha_I]-\frac{1}{4\pi}\alpha_I d \alpha_I +\frac{1}{2\pi}a d \alpha_I -\frac{1}{4\pi} ada
\end{equation}
where $\psi_I$ is a composite fermion from flux attachment to the boson $\varphi_I$. Here $I=1,2,3$. $\alpha_I$ is another emergent U(1) gauge field introduced to implement the flux attachment. We have a Fermi surface formed by the composite fermion:
\begin{align}
    \mathcal L_{\mathrm{FS}}[\psi_I,\alpha_I]&=\psi^*_I \left(i\partial_t+\alpha_{I;0}\right)\psi_I +\mu_\psi \psi^*_I \psi_I \notag \\ 
    &+\frac{\hbar^2}{2m^*}\sum_I \psi^*_I\left(-i\vec \nabla-\vec \alpha_I\right)^2 \psi_I
\end{align}
where the chemical potential $\mu_\psi$ fixes the density of each $\psi_I$ to be $\frac{x}{3}$ per triangular lattice unit cell.

So the final theory is:

\begin{align}
     \mathcal L&=\sum_{I=1,2,3}\big(\mathcal L_{FS}[\psi_I, \alpha_I]-\frac{1}{4\pi}\alpha_I d \alpha_I +\frac{1}{2\pi}a d\alpha_I\big) \notag \\ 
     &+\frac{3}{2\pi} A_c da +\frac{3}{4\pi} A_c d A_c
\end{align}

We can integrate $a$ to get: $\alpha_1=-A_c+\alpha, \alpha_2=-A_c+\beta, \alpha_3=-A_c-\alpha-\beta$.  In the end we have:
\begin{align}
     &\mathcal L_{\mathrm{holon-metal}}=-\frac{2}{4\pi}\alpha d \alpha-\frac{2}{4\pi}\beta d \beta-\frac{1}{2\pi} \alpha d \beta  \notag \\
     &+\mathcal L_{FS}[\psi_1, -A_c+\alpha]+\mathcal L_{FS}[\psi_2,-A_c+\beta] \notag \\ 
     &+\mathcal L_{FS}[\psi_3,-A_c-\alpha-\beta] \notag \\ 
     &-\frac{3}{4\pi}A_c d A_c+\frac{3}{4\pi} A_c d A_c
     \label{eq:holon_metal_CSL}
\end{align}
where in the second line we can see that the $\frac{3}{4\pi} A_c d A_c$ response from the stacking of the $C=3$ IQH phase is canceled.  

The translation symmetry acts as $T_x:\psi_I \rightarrow e^{i \frac{2\pi}{3} I} \psi_I, \alpha_\mu \rightarrow \alpha_\mu, \beta_\mu \rightarrow \beta_\mu$, $T_y: \psi_1 \rightarrow \psi_2, \psi_2 \rightarrow \psi_3, \psi_3 \rightarrow \psi_1, \alpha_\mu \rightarrow \beta_\mu, \beta_\mu \rightarrow -\alpha_\mu-\beta_\mu$.

\subsection{Property of the holon metal}

Here let us discuss the experimental property of the holon metal, with a focus on its transport signature.  We hope to calculate the conductivity tensor $\sigma_c=\begin{pmatrix} \sigma_{xx} & \sigma_{xy} \\ -\sigma_{xy} & \sigma_{yy} \end{pmatrix}$.
In the following, we assume that the translation symmetry is preserved.  Thus, $\psi_1, \psi_2, \psi_3$ have the same conductivity tensor $\sigma_{\psi_1}=\sigma_{\psi_2}=\sigma_{\psi_3}=\sigma_\psi$. It is easy to find the response to $A_c$ as:

\begin{align}
    \mathcal L_{\mathrm{eff}}&=\frac{1}{2} \sum_{\omega, \mathbf q}(A_{c;x}(\omega,q),A_{c;y}(\omega,q))3 \Pi(\omega, \mathbf q) \begin{pmatrix} A_{c;x}(-\omega,-\mathbf q)\\ A_{c;y}(-\omega,-\mathbf q) \end{pmatrix} \notag \\ 
    &~~~+\mathcal L[\alpha,\beta]
\end{align}
where we used the Coulomb gauge with $A_0=0$ and $\mathbf q \cdot \mathbf A(\omega, \mathbf q)=0$. There is no crossing term between $A_c$ and $\alpha,\beta$. $\Pi$ is the response from each $\psi_I$: $\Pi(\omega,\mathbf q=0)=\frac{e^2}{h} (-i\omega) \sigma_\psi= \frac{e^2}{h} (-i\omega)\begin{pmatrix} \sigma_{\psi;xx} & \sigma_{\psi;xy} \\ -\sigma_{\psi;-xy} & \sigma_{\psi;yy}\end{pmatrix}$.   So in the end, we have a simple relationship: $\sigma_c=3\sigma_\psi$, provided that the translation symmetry is not broken. Therefore, the holon metal has a resistivity tensor which is very similar to a Fermi liquid. Because the time-reversal symmetry is broken, $\psi$ can still have a small non-universal Hall conductivity. But the final phase has a small Hall angle $\frac{\rho_{xy}}{\rho_{xx}}$ because $\sigma_{xx}$ is large in the clean system.

\subsection{Alternative derivation and wavefunction}
To gain more insight into this exotic metallic phase, we also provide a different and perhaps simpler construction.   In this phase, there are three pockets with Fermi surface size $A_{\mathrm{FS}}=\frac{x}{3}$ per Brillouin zone (BZ).  We also find a gauge-invariant operator $C_{\mathrm{trion}}=\psi_1\psi_2\psi_3$ which is the physical trion operator. As discussed before, the physics is equivalent to doping a Laughlin state of the trion on top of a $C=3$ IQH phase.    Now the trion is at density $n_{\mathrm{trion}}=\frac{1}{3}-\frac{x}{3}$, we simply do parton construction $C_{\mathrm{trion}}=f_1 f_2 f_3$ and then put each flavor $f_I$ in a $C=-1$ Chern insulator plus a small hole pocket formed by $\psi_I\sim f_I^\dagger$ with size $x/3$.  In this construction, there is a SU(3) internal gauge field, but let us assume it's higgsed down to $U(1)\times U(1)$ by some terms. In the end we only have the U(1) gauge field $\alpha_\mu$ and $\beta_\mu$. $\psi_1$ couples to $-A_c+\alpha_\mu$,  $\psi_2$ couples to $-A_c+\beta_\mu$ and $\psi_3$ couples to $-A_c-\alpha_\mu-\beta_\mu$. The $C=-1$ Chern insulator gives a background Chern-Simons term. Eventually, we reproduce the theory in Eq.~\ref{eq:holon_metal_CSL} once we also include the $\frac{3}{4\pi} A_c d A_c$ term from the stacked $C=3$ IQH phase.

In this new approach, the resistivity tensor of the trion follows a generalized  Ioffe-Larkin rule, so the charge resistivity (without considering the $C=3$ IQH part) should be:
\begin{equation}
    \tilde \rho_c= \frac{1}{9}(\rho_1+\rho_2+\rho_3)
\end{equation}
where $\rho_I$ is the resistivity tensor from $f_I$. If the translation symmetry is unbroken, we have $\rho_1=\rho_2=\rho_3=\rho_f=\sigma^{-1}_f$.  Here $\sigma_f=\frac{e^2}{h}\begin{pmatrix} \sigma_{\psi;xx}  &-1+\sigma_{\psi;xy} \\ 1-\sigma_{\psi;xy} & \sigma_{\psi;yy}\end{pmatrix}=\sigma_\psi+\frac{e^2}{h}\begin{pmatrix} 0 & -1 \\ 1 & 0 \end{pmatrix}$.   Then we find $\tilde \sigma_c=3 \sigma_\psi+3\frac{e^2}{h}\begin{pmatrix} 0 & -1 \\ 1 & 0 \end{pmatrix} $. Note so far we haven't added the contribution from the $C=3$ IQH. The above is just the contribution from the trion sector.  The stacked $C=3$ IQH canceled the extra Hall conductivity, and we simply find $\sigma_c=\tilde \sigma_c+ 3 \frac{e^2}{h}\begin{pmatrix} 0 & 1 \\ -1 & 0 \end{pmatrix} =3\sigma_\psi$.  We need to emphasize that this simple relationship only holds when the translation symmetry is unbroken. If $\psi_1, \psi_2, \psi_3$ have different conductivity tensors, the final conductivity of the electron $\sigma_c$ has a more complicated expression: $\sigma_c=9(\rho_1+\rho_2+\rho_3)^{-1}+3 \frac{e^2}{h}\begin{pmatrix} 0 & 1 \\ -1 & 0 \end{pmatrix}$, where $\rho_I=\frac{h}{e^2}\begin{pmatrix} \sigma_{\psi_I;xx}  &-1+\sigma_{\psi_I;xy} \\ 1-\sigma_{\psi_I;xy} & \sigma_{\psi_I;yy}\end{pmatrix}^{-1} $.

\subsection{Pairing instability and charge-density-wave trion Fermi liquid}

The holon metal described in Eq.~\ref{eq:holon_metal_CSL} is actually unstable due to the gauge field fluctuations. In Eq.~\ref{eq:holon_metal_CSL} we ignored the Maxwell term for the internal gauge field $\alpha,\beta$, which is important for the propagator of the gauge field.  

We are interested in the $\omega \rightarrow 0$ limit. The self energy of $\alpha, \beta$ is mainly from the Landau diamagnetism of the Fermi pocket of $\psi_I$. This contribution leads to an additional term for the effective action of the internal gauge field:
\begin{align}
    \mathcal \delta L_{\mathrm{eff}}&=\frac{1}{2} \sum_{\omega, \mathbf q} \chi_d |q|^2 \big(|\alpha(\omega, \mathbf q)|^2  +|\beta(\omega,\mathbf q|^2 \notag \\ &+|\alpha(\omega,\mathbf q)+\beta(\omega,\mathbf q)|^2\big) \notag \\ 
    &=\chi_d\sum_{\omega,\mathbf q} q^2 (3|a_+(\omega,\mathbf q)|^2+|a_-(\omega,\mathbf q)|^2)
\end{align}
where $a_+=\frac{\alpha+\beta}{2}$ and $a_-=\frac{\alpha-\beta}{2}$.

The transverse gauge field then mediates repulsive or attractive interactions between the fermion $\psi_I$.  We find that $\psi_1$ couples to $a_++a_-$, $\psi_2$ couples to $a_+-a_-$ and $\psi_3$ couples to $-2 a_-$.  It is then easy to find that the intra-flavor BCS interaction $V_{\mathrm{intra}}(\mathbf q)\sim \frac{4 \upsilon_F^2}{3\chi_d q^2}$ and the inter-flavor BCS interaction $V_{\mathrm{inter}}(\mathbf q) \sim -\frac{2\upsilon_F^2}{3\chi_d q^2}$.  Therefore we find a strong attractive interaction between each pair of Fermi surfaces.

Because we have three Fermi surfaces, there is a frustration to pair all of them.  One natural state is to pair two of them and leave the third Fermi surface untouched. Hence we consider an ansatz with pairing $\delta H=\sum_{\mathbf k} \Delta(\mathbf k)\psi_1(\mathbf k) \psi_2(-\mathbf k)+h.c.$.  Now $\psi_1, \psi_2$ are gapped out, but $\psi_3$ is still gapless.  The state breaks the translation symmetry and triples the unit cell.  In our current notation the enlarged unit cell is a stripe with $1\times 3$ structure.  Note that we have $C^\dagger_{\mathrm{trion}}\sim \psi_1 \psi_2 \psi_3 \sim \langle \psi_1 \psi_2 \rangle \psi_3$, so after condensation of the pair $\langle \psi_1 \psi_2 \rangle \neq 0$ we can identify $\psi_3$ as the physical hole operator of the spinless trion.  Therefore, we get a Fermi liquid formed by the charge $3e$ trion, with the Fermi surface size equal to $\frac{x}{3}$ of the original Brillouin zone.

The final state depends on the angular momentum channel of the pairing $\Delta(\mathbf k)$.  Let us first discuss the simplest $s$-wave pairing. Then starting from the original theory in Eq.~\ref{eq:holon_metal_CSL}, we can get the low energy effective theory of the final CDW metal phase:

\begin{align}
     &\mathcal L_{\mathrm{CDW-metal}}=-\frac{2}{4\pi}\alpha d \alpha-\frac{2}{4\pi}\beta d \beta-\frac{1}{2\pi} \alpha d \beta  \notag \\
     &+\frac{1}{2\pi}\gamma_+d(-2A_c+\alpha+\beta) +\mathcal L_{FS}[\psi_3,-A_c-\alpha-\beta] 
\end{align}
where another internal U(1) gauge field $\gamma_+$ is introduced to describe the pairing $\psi_1 \psi_2$, which is charged under the gauge field $-2A_c+\alpha_\beta$.  The integration of $\gamma_+$ leads to $\alpha+\beta=2A_c$. Then $\psi_3$ couples to $-3A_c$, consistent with the expectation that $\psi_3$ is now a physical charge $-3e$ operator.  We are still left with the Chern-Simons terms. We can substitute $\alpha \rightarrow A_c+a$ and $\beta \rightarrow A_c-a$, then we reach the final theory:
\begin{align}
     &\mathcal L_{\mathrm{CDW-metal}}=\mathcal L_{FS}[\psi_3,-3A_c]-\frac{6}{4\pi}A_c d A_c -\frac{2}{4\pi} ad a \notag \\ 
\end{align}

It is interesting to see that this unconventional metallic phase has a Fermi surface formed by charge $-3e$ trion, coexisting with a Hall conductivity $\sigma_{xy}=-6 \frac{e^2}{h}$. There is also a hidden topological order, which is the same as the SU(2)$_1$ chiral spin liquid. We can also consider other pairing symmetry, but it only influences the hidden topological order and does not alter the transport property.

\section{Dope FCI:holon metal and CDW metal \label{sec:holon_metal_fci}}

We can now move to the discussion of doping a $\nu=-\frac{2}{3}$ FCI in the context of twisted MoTe$_2$. At $\nu=-1$, we have a Chern insulator with $C=1$. Then at $\nu=-\frac{2}{3}=-1+\frac{1}{3}$, we have an additional $1/3$ Laughlin state from electrons at $1/3$ filling of a $C=-1$ Chern band. This gives a Hall conductivity $\sigma_{xy}=\frac{2}{3} \frac{e^2}{h}$ at $\nu=-\frac{2}{3}$. Next we dope the FCI to the filling $\nu=-\frac{2}{3}-x=-1+\frac{1}{3}-x$. It can be understood as doping holes into the $\nu=-\frac{1}{3}$ Laughlin state and thus we expect similar physics as the doped SU(3)$_1$ CSL.

The doped anyons can be created by a boson field $\varphi^\dagger$, which is charged under an internal U(1) gauge field $a_\mu$ with self Chern-Simons term $\frac{3}{4\pi} a da$.  Again the projective translation symmetry guarantees that $\varphi$ has three minima in its dispersion. We label the low energy field as $\varphi_I$ with $I=1,2,3$. The effective theory is

\begin{align}
    \mathcal L&=\sum_{I=1,2,3} \varphi^*_I \left(i\partial_t+a_0\right)\varphi_I +\mu \varphi^*_I \varphi_I\notag \\ 
    &~~~+\frac{\hbar^2}{2m^*}\sum_I \varphi^*_I\left(-i\vec \nabla-\vec a\right)^2 \varphi_I \notag \\
    &~~~+\frac{3}{4\pi} a da+\frac{1}{2\pi}A da +\frac{1}{4\pi}A dA
    \label{eq:dope_FCI}
\end{align}
 $A$ is the probing field for the electron. One can see the similarity to Eq.~\ref{eq:dope_su3_csl}, but with the mapping $A_c=\frac{1}{3}A$.  We also replace the background term $\frac{3}{4\pi} A_c d A_c$ in Eq.~\ref{eq:dope_su3_csl} with $\frac{1}{4\pi} A dA$, which is from the $\nu=-1$ Chern insulator.

The action above is the standard effective theory of the Laughlin state. Each $\varphi_I$ creates an anyon with physical charge $-\frac{e}{3}$.  At physical hole density $x$, each $\varphi_I$ is at density $n_{\varphi_I}=x$. Variation of $a_0$ then leads to $\frac{\langle da \rangle}{2\pi}=-\frac{1}{3}\sum_{I=1,2,3}n_{\varphi_I}=-x$. So each $\varphi_I$ is at effective magnetic filling $\nu_m=-1$.

Again we need to deal with boson at $\nu_m=-1$. The translation symmetry acts as $T_x: \varphi_I \rightarrow e^{i \frac{2\pi}{3} I}\varphi_I$, $T_y: \varphi_1 \rightarrow \varphi_2, \varphi_2 \rightarrow \varphi_3$. Following the discussions in doped SU(3) CSL, one possible translation-invariant state is to put each $\varphi_I$ in the bosonic CFL phase.  We expect a similar holon metal phase as in Eq.~\ref{eq:holon_metal_CSL}, but with different charge assignment due to the $A_c=3A$ correspondence.  Let us label the composite fermion corresponding to each $\varphi_I$ as $\psi_I$, the low energy theory is:

\begin{align}
     \mathcal L&=\sum_{I=1,2,3}\big(\mathcal L_{FS}[\psi_I, \alpha_I]-\frac{1}{4\pi}\alpha_I d \alpha_I +\frac{1}{2\pi}a d\alpha_I\big) \notag \\ 
     &+\frac{1}{2\pi} A da +\frac{1}{4\pi} A d A
\end{align}

Integration of $a_\mu$ leads to the locking $\alpha_1=\alpha, \alpha_2=\beta, \alpha_3=-A-\alpha-\beta$. Then the final effective theory is:

\begin{align}
     &\mathcal L_{\mathrm{holon-metal}}=-\frac{2}{4\pi}\alpha d \alpha-\frac{2}{4\pi}\beta d \beta-\frac{1}{2\pi} \alpha d \beta  \notag \\
     &+\mathcal L_{FS}[\psi_1, \alpha]+\mathcal L_{FS}[\psi_2,\beta]+\mathcal L_{FS}[\psi_3, -A-\alpha-\beta] \notag \\ 
     &-\frac{1}{2\pi}Ad(\alpha+\beta)
     \label{eq:holon_metal_FCI}
\end{align}
Note here we use a different assignment of the physical charge from that in Eq.~\ref{eq:holon_metal_CSL}.

The phase is similar to the `secondary CFL' in Ref.~\cite{shi2024doping}, but there are differences in the coefficient of the self Chern-Simons term. The corresponding phase was dubbed as holon metal in the context of doped SU(3) CSL. Here we still call it holon metal just for simplicity. We note the property of the phase is very different from a CFL. Actually  it can also be constructed in a simpler way: we do the parton construction of the electron operator $c=f_1 f_2 f_3$. At $\nu=-1+\frac{1}{3}$, each $f_I$ is at density $n_f=\frac{1}{3}$. Then we let each $f_I$ form a Chern insulator with $C=-1$ with a tripled unit cell in the mean field level. This is the well-known construction of the Laughlin state. Then at hole doped case with $\nu=\frac{2}{3}-x$, each $f_I$ is at density $n_f=\frac{1}{3}-x$ and can form a small hole pocket with size $x$ on top of the $C=-1$ Chern insulator.  The resulting wavefunction is a product of three Slater determinants.  We can identify $\psi_I$ as the low energy field operator of the hole of $f_I$: $\psi_I \sim f_I^\dagger$.  One can also derive the same low energy field theory in this new construction (see Appendix.~\ref{sec:appendix}). Basically we can think in terms of the following simple picture: each hole is split to three fractionalized fermions with physical charge $-\frac{e}{3}$, which form three hole pockets on top of Chern insulators. Naively the mean field ansatz need to triple the unit cell. But the final phase is still translation invariant, just with a projective implementation: $T_x: \psi_I \rightarrow e^{i\frac{2\pi}{3}I}\psi_I$, $T_y: \psi_1 \rightarrow \psi_2, \psi_2 \rightarrow \psi_3, \psi_3 \rightarrow \psi_1$.

\subsection{Transport property}

The property of this exotic metallic phase is similar to the holon metal form doping the SU(3)$_1$ CSL. Just now $c \sim \psi_1^\dagger \psi_2^\dagger \psi_3^\dagger$ is a charge e hole instead of charge $3e$ trion.  Let us first discuss the transport properties. The conductivity tensor is: $\sigma_c=\tilde \sigma_c+ \frac{e^2}{h}\begin{pmatrix} 0 & 1 \\ -1 &0\end{pmatrix}$, where the second part is from the Chern insulator at $\nu=-1$.  $\tilde \sigma_c$ is the inverse of $\tilde \rho_c=\rho_1+\rho_2+\rho_3$, where $\rho_I$ is the resistivity tensor of the fermion $f_I$. Assuming translation symmetry, we have $\rho_1=\rho_2=\rho_3=\rho_f$ and $\tilde \sigma_c=\frac{1}{3} \frac{e^2}{h} \begin{pmatrix} \sigma_{\psi;xx} & \sigma_{\psi;xy}-1 \\ 1-\sigma_{\psi;xy} & \sigma_{\psi;xx}\end{pmatrix}$, where $\sigma_{\psi;xx}$ and $\sigma_{\psi;xy}$ are longitudinal and Hall conductivity in units of $\frac{e^2}{h}$ from the hole pocket $\psi_I$.  In the end, we have:

\begin{equation}
    \sigma_c=\frac{1}{3} \frac{e^2}{h}\begin{pmatrix} \sigma_{\psi;xx} & \sigma_{\psi;xy} \\ -\sigma_{\psi;xy} & \sigma_{\psi;xx}\end{pmatrix}+\frac{e^2}{h} \begin{pmatrix} 0 & \frac{2}{3} \\ -\frac{2}{3} & 0\end{pmatrix}
\end{equation}

Assuming $\sigma_{\psi;xy}$ is small given that $\psi$ does not feel any effective magnetic flux, we conclude that $\sigma_{c;xy}\approx \frac{2}{3} \frac{e^2}{h}$ in the holon metal phase. $\sigma_{c;xx}=\frac{1}{3} \sigma_{\psi;xx}$ depends on various details such as disorders. In the small $x$ limit we expect $\sigma_{\psi;xx} \sim x$ from the Drude model and is thus small. In this limit $\rho_{c;xy} \approx \frac{3}{2}\frac{h}{e^2}- A x$, and the Hall angle can be large. However, at larger $x$ we expect a large $\sigma_{\psi;xx}$ and thus $\rho_{c;xy}$ is significantly reduced from $\frac{3}{2} \frac{h}{e^2}$ with a small Hall angle.

\subsection{Instability to CDW metal}

Similar to the discussions for doped CSL, the holon metal here is also unstable due to the gauge field $\alpha_\mu,\beta_\mu$, which mediates inter-flavor attractive interaction. Again a natural ansatz has a pairing between two of the three pockets: $\delta H= \sum_{\mathbf k} \Delta(\mathbf k) \psi_1(\mathbf k) \psi_2(\mathbf k)+h.c.$.  The pairing breaks the translation symmetry and triples the unit cell. After the condensation $\langle \psi_1 \psi_2 \rangle \neq 0$, we have $c^\dagger \sim \psi_3$ and we can now identify $\psi_3$ as a physical hole operator. In the end we have a CDW metal with a single hole pocket formed by physical hole with charge $e$ now.

For simplicity, let us assume a simple s-wave pairing symmetry. Starting from Eq.~\ref{eq:holon_metal_FCI}, the pairing locks $\beta=-\alpha$. Then we get the following theory:

\begin{align}
     &\mathcal L_{\mathrm{CDW-metal}}=
     \mathcal L_{FS}[\psi_3, -A] -\frac{2}{4\pi}\alpha d \alpha
     \label{eq:holon_metal_FCI}
\end{align}

The first term just describes a hole pocket formed by the physical hole $\psi_3 \sim c^\dagger$.  The second term describes a hidden topological order, but it does not contribute to the physical response. We note that there is no background $AdA$ term, so we expect $\sigma_{c;xy} \approx 0$. This is different from a similar CDW metal in Ref.~\onlinecite{shi2024doping} where there the CDW metal coexists with an integer quantum Hall state. The difference origins from the different nature of the parent holon metal phase.

\subsection{Stabilization of the holon metal and quantum oscillations under magnetic field}

At zero magnetic field, we have shown that the holon metal is unstable to the CDW phase at lower temperature. In this subsection we demonstrate that the holon metal may be stabilized with a finite magnetic field $B$ and shows quantum oscillations.

Let us now apply magnetic field $B$ at a fixed doping $x$ for the holon metal.  Because each $\psi_I$ couples to internal gauge field $\alpha, \beta$, it is important to decide first how the internal flux respond to the external magnetic field $B$.  Let us define $\vec b_{\alpha}=\vec \nabla \times \vec \alpha$ and $\vec b_\beta=\vec \nabla \times \vec \beta$.  Starting from Eq.~\ref{eq:holon_metal_FCI}, simple variation of $\alpha_0, \beta_0$ leads to:

\begin{align}
    &\frac{2}{2\pi} b_{\alpha}+\frac{1}{2\pi}b_{\beta}=-\frac{1}{2\pi}B +n_1-n_3 \notag \\
    & \frac{2}{2\pi} b_{\beta}+\frac{1}{2\pi}b_{\alpha}=-\frac{1}{2\pi}B +n_2-n_3
\end{align}

Assuming translation symmetry, we have $n_1=n_2=n_3$ and therefore $b_{\alpha}=b_{\beta}=-\frac{1}{3}B$.  Hence there will be spontaneously generated internal flux such that each fermion $\psi_I$ feels an effective magnetic field of $-\frac{1}{3} B$, just like a particle with charge $-e/3$.

Because of the effective magnetic field, the inter-flavor pairing $\langle \psi_1 \psi_2 \rangle$ should be suppressed when $B$ is larger than a critical value. Therefore the holon metal may be the ground state at finite magnetic field.  Then we expect quantum oscillations in the form of $\cos \big( \frac{\hbar A_{\mathrm{FS}}}{e} \frac{1}{\frac{1}{3}B}-2\pi \gamma\big)$.   $\gamma$ is a phase shift from the background Berry phase.  Thus the quantum oscillation frequency is $f=\frac{3\hbar A_{\mathrm{FS}}}{e}$. Here $A_{\mathrm{FS}}$ is the Fermi surface volume corresponding to a density $x$ per original unit cell.  Note that this frequency is three times larger than a conventional CDW metal with hole pocket formed by charge $-e$ hole.

\section{Discussion on the experiment in twisted MoTe$_2$\label{sec:discussion}}

In our framework, the most natural symmetric non-quantum-Hall state upon doping the $\nu=-\frac{2}{3}$ FCI is a holon metal, which is then unstable to a CDW metal at lower temperature. On the other hand, we do not find a natural way to enter an `anyon superconductor' phase directly from an anyon gas, if the cheapest anyon is the elementary one with charge $e/3$. Nevertheless, signatures of superconductivity were reported from doping the $\nu=-\frac{2}{3}$ FCI in twisted MoTe$_2$\cite{xu2025signatures}. Here we briefly discuss possible scenarios, without touching the microscopic pairing mechanism.

\begin{figure}[ht]
    \centering
    \includegraphics[width=1\linewidth]{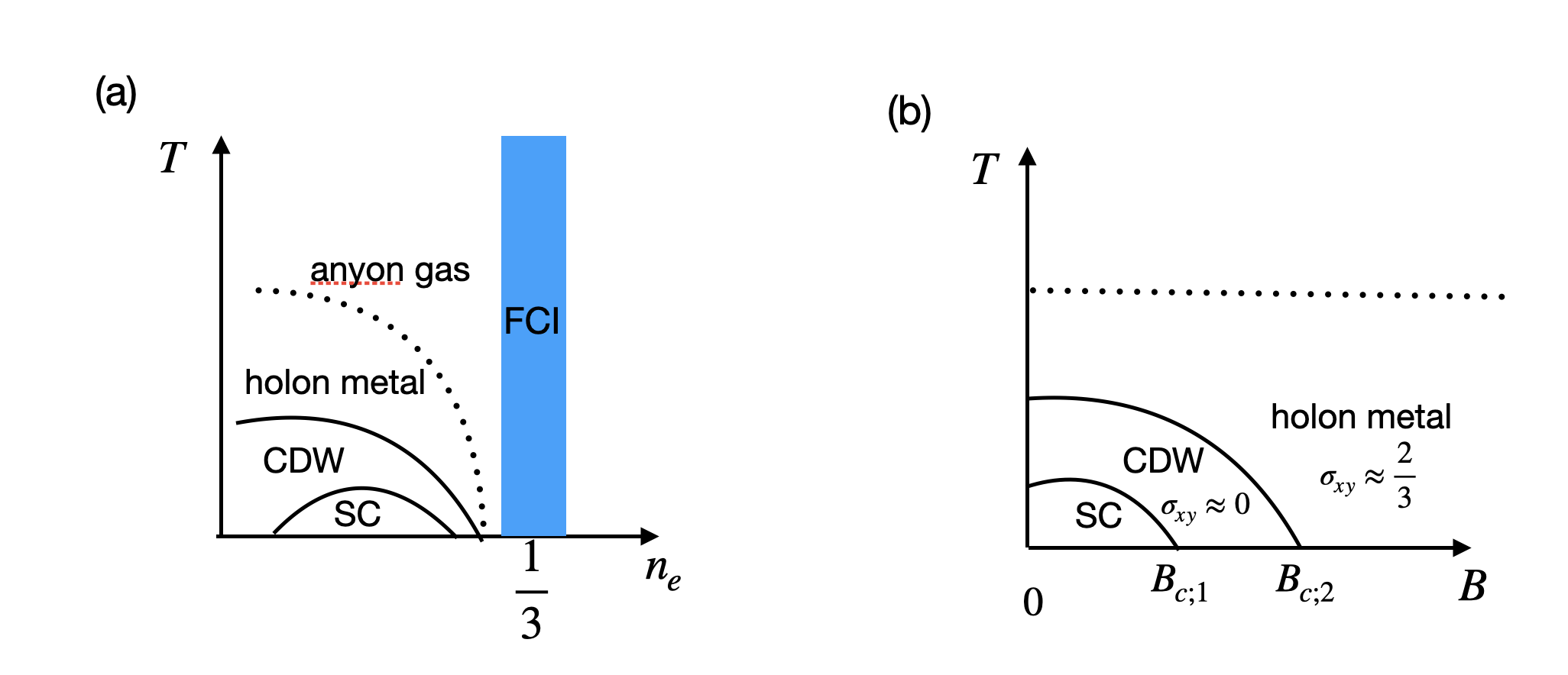}
    \caption{(a) Illustration of phase diagram with temperature and doping $n_e=\frac{1}{3}-x$ at small $x$. `SC' indicates superconductor and `CDW' indicates the CDW metal.  Anyon gas can just crossover to the holon metal without any sharp phase transition. The holon metal is unstable to the CDW metal. A superconductor, if exists, may likely emerge within the CDW metal phase, but the microscopic pairing mechanism is unclear. (b) At a fixed $x$, an illustrated phase diagram with the temperature $T$ and the magnetic field $B$.  In the holon metal phase, each pocket from $\psi_I$ feels an effective magnetic field of $-\frac{1}{3}B$, which can suppress the pairing and the instability into the CDW phase.}
    \label{fig:phase_diagram}
\end{figure}

(I) In the first  scenario, superconductivity emerges within the CDW metal. It will likely be in the $p\pm ip$ pairing.  The superconductor may or may not host additional deconfined anyons depending on the pairing symmetry of $\langle \psi_1(\mathbf k) \psi_2(-\mathbf k) \rangle$ (see detailed discussions in Appendix.~\ref{appendix:sc_cdw}).  A possible phase diagram is shown in Fig.~\ref{fig:phase_diagram}(a). The normal state just above $T_c$ is the CDW metal with coherent quasiparticle. The CDW phase enters the holon metal phase at a larger temperature, which cross overs to the anyon gas phase. We note that $\sigma_{xy} \approx 0$ in the CDW metal and $\sigma_{xy} \approx \frac{2}{3} \frac{e^2}{h}$ in the holon metal or anyon gas phase, so in this scenario we do not expect a jump of $\sigma_{xy}$ across $T_c$. A rapid increase of $\sigma_{xy}$ happens at a larger temperature. A similar evolution is also expected under the magnetic field. As shown in Fig.~\ref{fig:phase_diagram}(b), we expect an intermediate CDW metal phase when suppressing the superconductor by magnetic field $B$.  So at $T=0$, we expect a small $\sigma_{xy}$ just above $B_{c;1}$ where the superconductor is killed. A rapid increase of $\sigma_{xy}$ is expected at $B_{c;2}$ where the CDW metal transits to the holon metal phase.

(II) In the second  scenario, we can also imagine a  superonductor from intra-flavor $p+ip$ pairing $\langle \psi_I(\mathbf k) \psi_I(-\mathbf k) \rangle =\Delta(\mathbf k) \neq  0$ directly from the holon metal\cite{shi2024doping}.  The final state is topologically equivalent to three copies of $p+ip$ superconductors coexisting with an Abelian  topological order with $16$ anyons, as shown in Appendix.~\ref{eq:appendix_sc_2}. Note that intra-flavor pairing does not necessarily break the translation symmetry given that $T_y$ acts as an exchange of the flavor.  In this case we expect that the superconductor phase transits to the holon metal directly when increasing $T$ or $B$, so a rapid increase of $\sigma_{xy}$ is expected when the superconductor is killed. 

(III) In the more exotic scenario, one may imagine that the cheapest excitation of the FCI is the $2/3$ anyon for some unknown reasons. Although this is quite counterintuitive, there exist some numerical evidences to support it in fractional quantum Hall state\cite{xu2025dynamics}.   Ref.\cite{shi2024doping} showed that a superconductor equivalent to four copies of $p-ip$ superconductor can emerge in this scenario.  In Appendix.~\ref{appendix:sc_pairanyon}, we also explore this scenario in our framework and find a slightly more exotic superconductor. It also has chiral central charge of $c=-2$, but hosts a deconfined anti-semion.  In this exotic scenario, the normal state of the superconductor is very unconventional and is essentially a gas of $2/3$ anyons. 

 (IV) In the more conventional scenario, the superconductor is from a conventional Fermi liquid with an electron pocket of size $1/3-x$ and no symmetry breaking, as discussed in the weak coupling theory\cite{xu2025chiral,guerci2025fractionalization}. In this scenario, it is not clear how the Fermi liquid evolves to the FCI or anyon gas phase without an intermediate step. It is possible that the superconductivity instability is enhanced in a strongly correlated Fermi liquid proximate to the FCI phase\cite{wang2025chiral}. Ref.~\cite{zhang2025continuous} suggests that Fermi liquid phase close to the FCI may have an additional factor $|z_i-z_j|^6$ in its wavefunction. It is interesting to explore whether an effective attractive interaction arises for the quasi-particles in such a case.

 We leave it to the future experiments to measure the Hall number in the normal state, which can easily distinguish small hole pocket in our holon metal or CDW metal versus the large electron pocket in the conventional Fermi liquid.

\section{Conclusion \label{sec:conclusion}}

In summary, we provide a unified theoretical description of the doped fractional Chern insulator (FCI) at $\nu=-\frac{2}{3}$ and the doped SU(3)$_1$ chiral spin liquid. Assuming a finite spin gap, the doped SU(3)$_1$ system is dual to the doped FCI, where a spin-singlet charge-$3e$ trion corresponds to the electron in the FCI limit. In both regimes, a natural candidate phase is a holon metal featuring three small Fermi pockets. These pockets are formed by spinless holons carrying charge $-e$ in the doped CSL and charge $-e/3$ in the doped FCI. While this holon metal is likely unstable toward a charge density wave (CDW) state, it may be stabilized by an external magnetic field. We leave the investigation of the interplay between the holon metal and potential superconductivity to future work

\textbf{Acknowledgement} I thank Zhengyan Darius Shi for pointing out  the possibility of cheap charge $2e/3$ anyon. I thank Xue-Yang Song and Ashvin Vishwanath for earlier collaborations on doping SU(2) chiral spin liquid. The work is supported by the National Science Foundation under Grant No. DMR-2237031.

\bibliography{refs}

\appendix 

\onecolumngrid

\section{Transport property for holon metal from doping the FCI\label{sec:appendix}}

Let's provide an alternative derivation of the effective theory of the holon metal from hole doping the FCI at $\nu=-\frac{2}{3}$.  On top of the $\nu=-1$ Chern insulator, we have electron at density $n_e=\frac{1}{3}$. We then do the parton construction of the electron operator: $c(\mathbf r)=f_1(\mathbf r) f_2(\mathbf r) f_3(\mathbf r)$. This gives a SU(3) gauge redundancy, but we assume our mean field ansatz higgses it down to $U(1)\times U(1)$. There are two U(1) gauge fields $\alpha_\mu$ and $\beta_\mu$.  We also introduce the probing field $A_\mu$ and assign each $f_I$ to have charge $1/3$ under $A_\mu$.  Note this charge assignment can be shifted by redefining $\alpha$ and $\beta$.  As discussed in the main text, we can put each $f_I$ in a mean field ansatz with a tripled unit cell such that it can form a Chern insulator with $C=-1$ at filling $n_f=\frac{1}{3}$. Now at filling $n_f=\frac{1}{3}-x$, each $f_I$ just forms an additional hole pocket.  Let us first assume that each $f_I$ has an ansatz with uniform flux of $\frac{2\pi}{3}$ per triangular unit cell, so that there are three equivalent minima under hole doping. We label the hole-creation operator of these three minima as $\psi_{I;a}=f_I^\dagger$ with $I=1,2,3$ and $a=1,2,3$. Here $a$ is the valley index that labels the three minima.   In principle, each $f_I$ can form three hole pockets, occupying all three valleys. However, we can reach a simpler ansatz by adding perturbations which split the energy of these three valleys so that $f_I$ forms a single hole pocket corresponding to $a=I$. Naively, this breaks the translation symmetry. However, we can always combine an exchange between the three $f_I$ in translation symmetry. More precisely, we have the following implementation of the translation symmetry: $T_x: \psi_{I;a} \rightarrow e^{i\frac{2\pi}{3}I}\psi_{I;a}$ and $T_y: \psi_{I;a} \rightarrow \psi_{I+1;a+1}$. Here $I+1$ and $a+1$ are defined as mod $3$ so $3+1=1$. One can check that a perturbation term like $\delta H=-\Delta \sum_{I=1,2,3}\psi^\dagger_{I;I}\psi_{I;I}$ is allowed.  Then in the low energy we have one single hole pocket from each $f_I$, formed by the low energy field operator $\psi_I=\psi_{I;I}$.

This approach also enables us to write down a variational wavefunction for the holon metal:

\begin{equation}
    \psi_{\mathrm{holon-metal}}(z_1,z_2,...,z_N)=\prod_{I=1,2,3}\text{Slater}_{f_I}(z_1,...,z_N)
\end{equation}
where $z_i$ is the coordinate of the electron on top of the $\nu=-1$ insulator.  Here we have a product of three Slater determinant. Each Slater determinant is the wavefunction for $f_I$, which is captured by a mean field ansatz with $\frac{2\pi}{3}$ flux per original unit cell.

Next we derive the low energy effective theory. In our theory $f_1$ couples to $\frac{1}{3}A_\mu-\alpha_\mu$, $f_2$ couples to $\frac{1}{3}A_\mu-\beta_\mu$ and $f_3$ couples to $\frac{1}{3}A_\mu+\alpha_\mu+\beta_\mu$.  $c=f_1f_2f_3$ is gauge invariant and only couples to $A_\mu$.  Because each $f_I$ has a background Chern insulator, we get a background Chern-Simons term: $-\frac{1}{4\pi}(\frac{1}{3}A-\alpha) d(\frac{1}{3}A-\alpha)-\frac{1}{4\pi}(\frac{1}{3}A-\beta)d(\frac{1}{3}A-\beta)-\frac{1}{4\pi}(\frac{1}{3}A+\alpha+\beta)d(\frac{1}{3}A+\alpha+\beta)$. The final theory is:

\begin{align}
     &\mathcal L_{\mathrm{holon-metal}}=-\frac{2}{4\pi}\alpha d \alpha-\frac{2}{4\pi}\beta d \beta-\frac{1}{2\pi} \alpha d \beta +\frac{2}{3} \frac{1}{4\pi} A dA \notag \\
     &+\mathcal L_{FS}[\psi_1, -\frac{1}{3}A+\alpha]+\mathcal L_{FS}[\psi_2,-\frac{1}{3}A+\beta]+\mathcal L_{FS}[\psi_3, -\frac{1}{3}A-\alpha-\beta] \notag \\ 
\end{align}
Note that we also added a background term $\frac{1}{4\pi} A d A $ from the $\nu=-1$ Chern insulator.

In the above we have a fractional Chern-Simons term for $A$. We can redefine $\alpha \rightarrow \alpha+\frac{1}{3} A$ and $\beta \rightarrow \beta+\frac{1}{3} A$. Then we recover Eq.~\ref{eq:holon_metal_FCI} in the main text.  The current form also matches Eq.~\ref{eq:holon_metal_CSL} using the correspondence $A_c=\frac{1}{3}A$ if we add the additional background term $\frac{2}{3} \frac{1}{4\pi} A dA$.

\section{Topological property of superconductor from CDW metal\label{appendix:sc_cdw}}

Here we discuss SC from CDW metal. CDW metal is understood as from inter-flavor pairing $\langle \psi_1 (\mathbf k) \psi_2(-\mathbf k) \rangle =\Delta_{12}(\mathbf k)$ in the holon metal. Depending on the pairing symmetry of $\Delta_{12}$, it may coexists with an Abelian topological order.  As a result, the final superconductor from pairing $\langle \psi_3(\mathbf k) \psi_3(-\mathbf k) \rangle =\Delta_{33}(\mathbf k)$ may coexits with a topological order. We assume $\Delta_{33}(\mathbf k)$ is in $p\pm ip$ pairing. In the following we discuss different possibilities depending on the angular momentum of $\Delta_{12}(\mathbf k)$.

\subsection{s-wave $\Delta_{12}(\mathbf k)$}

In this case we have a $p\pm ip$ SC coexisting with $U(1)_2$ topological order. The chiral central charge is $c=1\pm \frac{1}{2}$.  Starting from the theory for the holon metal in Eq.~\ref{eq:holon_metal_FCI}, a non-zero s wave $\Delta_{12}$ can be described by a simple term $-\frac{1}{2\pi} a d(\alpha+\beta)$, which then locks $\beta=-\alpha$. In the end the CDW metal is simply a Fermi liquid of $\psi_3$ plus a $U(1)_2$ topological order described by $\frac{2}{4\pi}\alpha d \alpha$. Note that this  topological order does not couple to $A$ and only contributes a chiral central charge of $c=1$.  If we further introduce $p\pm ip$ pairing for $\psi_3$, we simply have a $p\pm ip$ superconductor coexisting with a hidden topological order with $2$ anyons.

\subsection{$\Delta_{12}$ in $p+ip$ channel}

We assume a $p+ip$ pairing for $\Delta_{12}$.  And also a $p\pm ip$ pairing for $\Delta_{33}$.   $\Delta_{12}$ gives a term $-\frac{1}{4\pi}(a_2-a_3)d(a_2-a_3)-\frac{1}{2\pi}a_2d\alpha-\frac{1}{2\pi}a_3 d \beta$, where $a_2, a_3$ are introduced to represent the dual field of the pairing $\Delta_{12}$ and also the topological response.  $\Delta_{33}$ gives a term $\frac{2}{2\pi}a_1 d (-A-\alpha-\beta)$.  One should keep in mind that the unit charge of $a_1$ binds a Majorana zero mode.

In the basis of $a=(a_1, a_2, a_3, \alpha,\beta)$ we can write down the final theory for the superconductor as:
\begin{equation}
    \mathcal L_{\mathrm{SC}}=-\frac{1}{4\pi} a^T K a -\frac{1}{2\pi} A q^T d a
\end{equation}
with $q=(2,0,0,1,1)^T$ and

\begin{equation}
    K=\left(
\begin{array}{ccccc}
 0 & 0 & 0 & 2 & 2 \\
 0 & 1 & -1 & 1 & 0 \\
 0 & -1 & 1 & 0 & 1 \\
 2 & 1 & 0 & 2 & 1 \\
 2 & 0 & 1 & 1 & 2 \\
\end{array}
\right)
\end{equation}

By doing a transformation $a=W\tilde a$ with $W=\left(
\begin{array}{ccccc}
 -1 & 0 & 0 & 0 & 0 \\
 2 & 1 & 0 & 0 & 0 \\
 2 & 0 & 1 & 0 & 0 \\
 0 & 0 & 0 & 1 & 0 \\
 0 & 0 & 0 & 0 & 1 \\
\end{array}
\right)$, we can get a new $K$ matrix $\tilde K=\begin{pmatrix} 0 & 0 \\ 0 & K' \end{pmatrix}$ with $K'$ a $4\times 4$ matrix:
\begin{equation}
    K'=\left(
\begin{array}{cccc}
 1 & -1 & 1 & 0 \\
 -1 & 1 & 0 & 1 \\
 1 & 0 & 2 & 1 \\
 0 & 1 & 1 & 2 \\
\end{array}
\right)
\end{equation}

Now the charge vector is $\tilde q=(-2,0,0,1,1)^T$.  One can check that $\det W=1$ so  the charge quantization for $\tilde a$ is not altered. We label the first gauge field for $\tilde K$ as $\tilde a_1$, then there is a term $-\frac{2}{2\pi}A d \tilde a_1$, indicating a charge $2e$ superconductor, which coexists with a decoupled sector described by $K'$. One can check that the unit charge of $a_1$ also carries unit charge of $\tilde a_1$ and vice versa. Hence the unit charge of $\tilde a_1$, the elementary vortex of the superconductor, binds a Majorana zero mode.  Meanwhile  $|\det K'|=1$, so there is no deconfined anyon.   The final phase is basically just a conventional $p\pm ip$ superconductor.  But there is a chiral central charge of $2$ from $K'$, so the final central charge is $c=3\pm \frac{1}{2}$ for $p\pm ip$ pairing of $\psi_3$.

\subsection{$\Delta_{12}$ in p-ip pairing}

We can repeat the same analysis as the previous subsection. Just now the K matrix has an extra minus sign in the block of $(a_2,a_3)$. We have

\begin{equation}
    K=\left(
\begin{array}{ccccc}
 0 & 0 & 0 & 2 & 2 \\
 0 & -1 & 1 & 1 & 0 \\
 0 & 1 & -1 & 0 & 1 \\
 2 & 1 & 0 & 2 & 1 \\
 2 & 0 & 1 & 1 & 2 \\
\end{array}
\right)
\end{equation}
and $q=(2,0,0,1,1)^T$.

Again we need a transformation with the same $W=\left(
\begin{array}{ccccc}
 -1 & 0 & 0 & 0 & 0 \\
 2 & 1 & 0 & 0 & 0 \\
 2 & 0 & 1 & 0 & 0 \\
 0 & 0 & 0 & 1 & 0 \\
 0 & 0 & 0 & 0 & 1 \\
\end{array}
\right)$ to reach $\tilde K=\begin{pmatrix} 0 & 0 \\ 0 & K' \end{pmatrix}$ with $K'$ a $4\times 4$ matrix:
\begin{equation}
    K'=
\left(
\begin{array}{cccc}
 -1 & 1 & 1 & 0 \\
 1 & -1 & 0 & 1 \\
 1 & 0 & 2 & 1 \\
 0 & 1 & 1 & 2 \\
\end{array}
\right)
\end{equation}
and $\tilde q=(2,0,0,1,1)^T$.

Now one can check that $\det K'=3$ with zero chiral central charge.  $K'$ describes a topological order similar to the $U(1)_{3}$ state stacked with integer quantum Hall phase.  In the end we have a $p\pm ip$ superconductor coexisting with a topological order with $3$ anyons. The chiral central charge is $c=\pm \frac{1}{2}$, purely from the $p\pm ip$ pairing of $\Delta_{33}$.

\section{Superconductor directly from the holon metal\label{eq:appendix_sc_2}}

We discuss the topological property of the exotic superconductor from $p+ip$ intra-flalvor pairing of the holon metal in Eq.~\ref{eq:holon_metal_FCI}. We introduce the dual gauge field $a_1, a_2, a_3$ to encode the pairing for the three pockets $\langle \psi_I \psi_I \rangle \neq 0$.  The final low energy effective theory is:

\begin{align}
     &\mathcal L_{\mathrm{SC}}=-\frac{2}{4\pi}\alpha d \alpha-\frac{2}{4\pi}\beta d \beta-\frac{1}{2\pi} \alpha d \beta 
     +\frac{2}{2\pi}a_1 d\alpha+\frac{2}{2\pi} a_2 d\beta+\frac{2}{2\pi}a_3d(-A-\alpha-\beta)-\frac{1}{2\pi}Ad(\alpha+\beta) \notag \\ 
    & =-\frac{1}{4\pi} a^T K a -\frac{1}{2\pi} A q^T d a
\end{align}
where in the second line we use the notation that $a=(a_1,a_2,a_3,\alpha,\beta)^T$. We have $q=(0,0,2,1,1)^T$ and

\begin{equation}
K=\left(
\begin{array}{ccccc}
 0 & 0 & 0 & -2 & 0 \\
 0 & 0 & 0 & 0 & -2 \\
 0 & 0 & 0 & 2 & 2 \\
 -2 & 0 & 2 & 2 & 1 \\
 0 & -2 & 2 & 1 & 2 \\
\end{array}
\right)    
\end{equation}
one can find $\det K=0$, indicating a gapless Goldstone mode. Also the excitation can be labeled by $l=(l_1,l_2,l_3,l_4,l_5)^T$ under $a$. Note that we need to bind a Majorana mode to the vortex of the pairing $\langle \psi_I \psi_I \rangle$, which is represented by the unit charge of $a_1,a_2, a_3$. Therefore $V_{\mathrm{Ising}}=l_1+l_2+l_3$ labels the number of Majorana mode bound to the excitation mod $2$.

We make a transformation such that $a \rightarrow W a$, $K \rightarrow W^T K W$ and $q \rightarrow W^T q$. Similarly any excitation with charge vector $l$ under $a$ transforms as: $l \rightarrow W^T l$.  We choose 

\begin{equation}
    W=\left(
\begin{array}{ccccc}
 1 & 0 & -1 & 0 & 0 \\
 1 & 1 & -1 & 0 & 0 \\
 1 & 0 & 0 & 0 & 0 \\
 0 & 0 & 0 & 1 & 0 \\
 0 & 0 & 0 & 0 & 1 \\
\end{array}
\right)
\end{equation}
with $\det W=1$ satisfied. This leads to a new K matrix as:

\begin{equation}
    \tilde K=\left(
\begin{array}{ccccc}
 0 & 0 & 0 & 0 & 0 \\
 0 & 0 & 0 & 0 & -2 \\
 0 & 0 & 0 & 2 & 2 \\
 0 & 0 & 2 & 2 & 1 \\
 0 & -2 & 2 & 1 & 2 \\
\end{array}
\right)
\end{equation}
with $\tilde q=(2,0,0,1,1)^T$.

Let us label the first gauge field as $\gamma$ and the other four gauge field as $\beta=(\beta_1,\beta_2, \beta_3, \beta_4)^T$. So the final action is 

\begin{equation}
    \mathcal L_{\mathrm{SC}}=-\frac{2}{2\pi} A d \gamma- \frac{1}{2\pi}Ad(\beta_3+\beta_4)-\frac{1}{4\pi} \beta^T  K' \beta
\end{equation}
with $K'=\left(
\begin{array}{cccc}
 0 & 0 & 0 & -2 \\
 0 & 0 & 2 & 2 \\
 0 & 2 & 2 & 1 \\
 -2 & 2 & 1 & 2 \\
\end{array}
\right)$.

One also find that in the new basis, the charge under $\gamma$ is  $\tilde l_\gamma=l_1+l_2+l_3=V_{\mathrm{Ising}}$. The unit charge of $\gamma$ represents the vortex of this charge $2e$ superconductor. Thefore for each vorex hosts a Majorana mode mod $2$ and single Majorana mode only exists in the superconductor vortex. 
 We have $\det K'=16$ and the associated chiral central charge is $0$. The three $p+ip$ pairings provide a net chiral central charge of $c=3$. Therefore  the final phase is topologically equivalent to three copies of $p+ip$ superconductor stacked with an Abelian topological order with $16$ anyons.  We note that the $\frac{1}{2\pi} A d(\beta_3+\beta_4)$ term indicates that the associated Abelian order has anyons with fractional charge.  The minimal charge is carried by an anti-semion with charge $e/2$.

\section{Superconductor when doping $Q=\frac{2}{3}e$ anyons \label{appendix:sc_pairanyon} }

If the cheapest excitation of  the FCI is a pair of $1/3$ anyons, there is a direct way to enter the superconductor. We do not have a microscopic understanding why this can happen energetically. In the following we assume that the $2/3$ anyon is cheaper and just discuss the consequences. In this case, the low energy effective theory should be dominated by a boson $\Phi^\dagger$, which creates the $2/3$ anyon. Note that $\Phi$ feels a background $\frac{4\pi}{3}$ flux, so there are still three minima in its dispersion and we have three flavors labeled as $\Phi_I$, with $I=1,2,3$. The low energy theory is:

\begin{align}
    \mathcal L&=\sum_{I=1,2,3} \Phi^*_I \left(i\partial_t+2a_0\right)\Phi_I +\mu \Phi^*_I \Phi_I+\frac{\hbar^2}{2m^*}\sum_I \Phi^*_I\left(-i\vec \nabla-2\vec a\right)^2 \Phi_I \notag \\
    &~~~+\frac{3}{4\pi} a da+\frac{1}{2\pi}A da +\frac{1}{4\pi}A dA
    \label{eq:dope_FCI_pair_anyon}
\end{align}
 Note that $\Phi_I$ carries charge $2$ under $a_\mu$, so it creates a $2/3$ anyon.  At physical density $n=-\frac{2}{3}-x$, we have $n_{\Phi_I}=\frac{x}{2}$. From the variation of $a_0$ we find that $\frac{\langle da \rangle}{2\pi}=-\frac{2}{3} \sum_I n_{\Phi_I}=-x$.  Each $\Phi_I$ is thus at effective magnetic filling of $\nu_m=-\frac{1}{4}$. We can simply put each $\Phi_I$ in a $-1/4$ Laughlin state.

 In the end, we get a theory of superconductor

\begin{align}
    \mathcal L&=\sum_{I=1,2,3}\frac{4}{4\pi}\beta_I d \beta_I+\frac{2}{2\pi} a d \beta_I
    +\frac{3}{4\pi} a da+\frac{1}{2\pi}A da +\frac{1}{4\pi}A dA 
    \label{eq:sc_pair_anyon}
\end{align}

We can make a redefinition: $a=-2\tilde a, \beta_I=\tilde a+\tilde \beta_I$, so the theory is simply

\begin{equation}
    \mathcal L=-\frac{2}{2\pi}A d\tilde a+\sum_I \frac{4}{4\pi}\tilde \beta_I d \tilde \beta_I+\frac{1}{4\pi} A dA
\end{equation}

The charge of the new gauge field is related to the old ones by $q_{\tilde a}=-2q_a+\sum_I q_{\beta_I}$ and $q_{\tilde \beta_I}=q_{\beta_I}$.  Define $\tilde l=(q_{\tilde a},q_{\tilde \beta_1},q_{\tilde \beta_2},q_{\tilde \beta_3})$ and $l=(q_a,q_{\beta_1},q_{\beta_2},q_{\beta_3})$. One can see that $\tilde l=(1,1,0,0)$ is allowed by $l=(0,1,0,0)$.  Therefore the minimal charge of $\tilde a$ is $1$ and this creates the elementary vortex of the charge $2e$ superconductor. It has a statistics of $\theta=-\frac{\pi}{4}$. Because it carries a vortex charge, it costs infinite energy and should not be identified as a deconfined anyon.  There is an elementary anyon with $\tilde l=(0,1,1,0)$, corresponding to $l=(1,1,1,0)$. It has statistics of $\theta=-\frac{\pi}{2}$ and is an anti-semion. Including the trivial anyon, there are in total two deconfined anyons.   The chiral central charge of this superconductor is $c=-2$ if we also include the background integer quantum Hall part from the $\nu=-1$ state.

\end{document}